\documentclass[11pt,notoc]{JHEP3}

\usepackage{epsfig}

\usepackage{amsmath}
\usepackage{mathrsfs}
\usepackage{amsmath,amssymb}
 \usepackage{graphicx}

\def\bZ {\bb{Z}}

\def\bE {\bb{E}}

\makeatletter
\@addtoreset{equation}{section}
\makeatother

\def\ben{\begin{equation}}
\def\een{\end{equation}}

  \let\n=\nu  \let\p=\pi

\let\C=\Chi

 \def\bd{\begin{document}} \def\ed{\end{document}}
\def\ds{\documentstyle} \let\fr=\frac \let\bl=\bigl \let\br=\bigr
\let\Br=\Bigr \let\Bl=\Bigl
\let\bm=\bibitem
\let\na=\nabla
\let\pa=\partial \let\ov=\overline
\newcommand{\be}{\begin{equation}}
\newcommand{\ee}{\end{equation}}
\def\ba{\begin{array}}
\def\ea{\end{array}}
\def\ft#1#2{{\textstyle{\frac{\scriptstyle #1}{\scriptstyle #2} } }}
\def\fft#1#2{{\frac{#1}{#2}}}
\def\del{\partial}
\def\vp{\varphi}
\def\sst#1{{\scriptscriptstyle #1}}
\def\oneone{\rlap 1\mkern4mu{\rm l}}
\def\td{\tilde}
\def\wtd{\widetilde}
\def\ie{{\it i.e.\ }}
\def\dalemb#1#2{{\vbox{\hrule height .#2pt
        \hbox{\vrule width.#2pt height#1pt \kern#1pt
                \vrule width.#2pt}
        \hrule height.#2pt}}}
\def\square{\mathord{\dalemb{6.8}{7}\hbox{\hskip1pt}}}
\newcommand{\ho}[1]{$\, ^{#1}$}
\newcommand{\hoch}[1]{$\, ^{#1}$}
\newcommand{\bea}{\begin{eqnarray}}
\newcommand{\eea}{\end{eqnarray}}
\newcommand{\ra}{\rightarrow}
\newcommand{\lra}{\longrightarrow}
\newcommand{\Lra}{\Leftrightarrow}
\newcommand{\bp}{\tilde \beta^\prime}
\newcommand{\tr}{{\rm tr} }
\newcommand{\Tr}{{\rm Tr} }

\def\LL{{\cal L}}
\def\TT{{\cal T}}
\def\NN{{\cal N}}
\def\VV{{\cal V}}
\def\p{{\partial}}
\def\ov{\over}
\def\apr{{\alpha'}}

\def\0{{\sst{(0)}}}
\def\1{{\sst{(1)}}}
\def\2{{\sst{(2)}}}
\def\3{{\sst{(3)}}}
\def\4{{\sst{(4)}}}
\def\5{{\sst{(5)}}}
\def\6{{\sst{(6)}}}
\def\7{{\sst{(7)}}}
\def\8{{\sst{(8)}}}
\def\n{{\sst{(n)}}}
\def\cA{{{\cal A}}}
\def\cB{{{\cal B}}}
\def\cF{{{\cal F}}}
\def\cG{{{\cal G}}}
\def\cH{{{\cal H}}}
\def\tV{\widetilde V}
\def\tW{\widetilde W}
\def\tH{\widetilde H}
\def\tE{\widetilde E}
\def\tF{\widetilde F}
\def\tA{\widetilde A}
\def\im{{{\rm i}}}
\def\tY{{{\wtd Y}}}
\def\ep{{\epsilon}}
\def\vep{{\varepsilon}}
\def\bD{{{\bar D}}}
\def\R{{{\mathbb R}}}
\def\C{{{\mathbb C}}}
\def\H{{{\mathbb H}}}
\def\CP{{{\mathbb C}{\mathbb P}}}
\def\RP{{{\mathbb R}{\mathbb P}}}
\def\Z{{{\mathbb Z}}}
\def\bA{{{\mathbb A}}}
\def\bB{{{\mathbb B}}}
\def\bC{{{\mathbb C}}}
\def\bD{{{\mathbb D}}}
\def\bE{{{\mathbb E}}}
\def\bZ{{{\mathbb Z}}}
\def\Re{{{\frak{Re}}}}
\def\Im{{{\frak{Im}}}}
\def\cosec{{\,\hbox{cosec}\,}}
\def\Gm{{\Gamma_{\!\! -}}}
\def\Gp{{\Gamma_{\!\! +}}}
\def\stan{{standard }}
\def\nonstan{{supernumerary }}
\def\p{{\partial}}
\def\kdel#1{{\fft{\del}{\del#1}}}

\def\bog{{Bogomolny }}
\def\om{{\omega}}

\newcommand{\icts}{\it Interdisciplinary Center of Theoretical
Studies, USTC, Hefei, Anhui 230026, PRC}

\newcommand{\itp}{\it Institute of Theoretical Physics, CAS, Beijing 100190, PRC}

\newcommand{\auth}{
 Yang Zhou\hoch{\dagger\ddagger}}

\title{D4 brane probes in gauge/gravity duality}

\author{Yang Zhou\\
Institute of Theoretical Physics\\
Chinese Academic of Science\\
Beijing 100190, PRC\\

Interdisciplinary Center of Theoretical Studies\\
USTC, Hefei, \\
Anhui 230026, PRC \\}

\abstract{We propose a DBI vertex brane + $N_c$ fundamental strings
configuration for a probe baryon in the finite-temperature thermal
gauge field via AdS/CFT correspondence. In particular, we
investigate properties of this configuration in QCD$_4$ and warped
AdS$_6\times $S$^4$. We find that, in D4-D8 system, a holographic
probe baryon can be described as N$_c$ fundamental strings
connecting through a vertex D4 brane wrapped on S$^4$. In QCD$_4$
background, a closed vertex can exist in confined phase and can not
exist in deconfined phase. In the low temperature region, screening
effect still exist in confined phase like meson and the vertex D4
brane dominates the baryon mass. The lower energy state corresponds
to vertex brane closer to the radial cut off position ($r=r_c$) and
the higher energy state corresponds to vertex brane a little far
away from the cut off position. The high energy limit of this
configuration is just like the unclosed vertex brane configuration
in a higher temperature deconfined phase. In warped
AdS$_6\times$S$^4$ background, a closed vertex can exist in
deconfined phase and the vertex contains a spike, while fundamental
strings are relatively short. Screening length should be defined
through the distance between top position of the vertex spike and
the boundary.}

\keywords{D-branes, thermal field theory}

\begin{document}


\thispagestyle{empty}

\pagebreak
%
%


\tableofcontents

\section{Introduction}

In recent years, there has been much interest in studying hadrons in strongly coupled QCD in terms of AdS/CFT
correspondence. One important topic is the study of holographic baryons in gravity
background~\cite{GI0801,SS0802,Ghoroku:2008na,Sfetsos:2008yr}. A holographic baryon in gauge/gravity duality was first
introduced by Witten~\cite{Witten:1998}, where a baryon is identified with a compact D-brane wrapped on a transverse
sphere with $N_c$ strings attached to it.
Investigations of baryons in AdS/CFT have been started since ten
years ago~\cite{CGST:9902,CGST:9810}, in order to find a solution
with a compact vertex brane and DBI strings, but there are many
challenges~\cite{JLR:0606,HP:0405,Ima:0410,CGST:9902,CGST:9810}. One
big problem is how to get a closed brane solution for baryon vertex
from the DBI+CS action in a certain gravity background which is dual
to the gauge field we want to study. We construct a new
configuration with a wrapped vertex brane and N$_c$ strings to solve
these problems. Furthermore, we investigate properties of this
configuration in probe limit and argue that these properties may
have some signals in QGP in RHIC experiment.

Recently, a simple configuration of baryon in a hot strongly coupled Super Yang Mills plasma was
proposed\cite{Liu:2008bs}, where the screening length of a moving baryon with finite velocity was investigated.
Screening length and J-E$^2$ behavior(J is angular momentum and E is baryon mass) of high spin baryons were
analyzed~\cite{Li:2008py}. All these investigations are in the framework of thermal SYM gauge theory/AdS black hole
duality and component quarks are considered as probes. The main results of these works show that baryons in a hot
strongly coupled plasma have screening length, which is similar to meson case. Boost velocity and angular velocity
dependence of screening length for baryons are both similar to those of mesons. These results are natural and
reasonable, because there the vertex brane is treated as a massive point in AdS$_5$, with an action only depending on
the gravity potential.

Generally, the vertex brane can be not a point in AdS$_5$, but a line(trajectory of wrapped brane on S$^5$), which has
an action with a DBI term and a Chern-Simons term. Though it's difficult to find a good solution for the vertex brane
in many gravity backgrounds, in the new configuration we propose, we obtain two kinds of solutions in QCD$_4$ and
AdS$_6\times$S$^4$ background respectively, each of which has one closed vertex brane and N$_c$ hanging
strings.\footnote{QCD$_4$ has the same mean as one in paper ~\cite{CGST:9902}.} From new configurations constructed, we
find that baryons have some special properties in the probe limit. An apparent property is that baryon mass may be
dominated by the vertex brane, so the "melting picture" of probe baryon in the quark gluon plasma is very different
from the meson melting.


Another way to see the role of baryon vertex in gauge/gravity framework is through the finite quark density(or we also
say baryon density) D$_\textrm{color}$-D$_\textrm{flavor}$ system. Chemical potential of finite quark density was
introduced as time component of U(1) gauge field on flavor branes~\cite{Kobayashi:2006sb}, and chemical potential of
isospin density was introduced as time component of SU(2) gauge field on flavor branes. Finite quark and isospin
density affect embedding of flavor brane, as well as the phase transition corresponding to meson
dissociation~\cite{Erdmenger:2007ja,Mas:2008jz,Erdmenger:2008yj}. In a D$_\textrm{color}$-D$_\textrm{flavor}$ system ,
a flavor brane can have Minkowski embedding or black hole embedding. But in the case of finite quark density, it has
been argued that D$_\textrm{flavor}$ branes have to touch the black hole horizon since the strings connecting the
D$_\textrm{flavor}$ branes and the horizon can be replaced by the deformation of the D$_\textrm{flavor}$
brane~\cite{Kobayashi:2006sb}, thus there is no Minkowski embedding in the finite quark density case. In this case, a
baryon vertex can end hanging strings outside of the horizon, so there is no way for a flavor brane to touch the
horizon~\cite{SS0802}. Since this argument, we can see that we can safely discuss a baryon in the probe limit in the
Minkowski embedding. We should also note that a baryon in our probe limit is different from the baryons in finite
density which should be considered as background baryons. In our probe limit, a baryon almost have no backreaction on
flavor branes.

To find a suitable D$_\textrm{color}$-D$_\textrm{flavor}$ system for discussion, we notice the QCD$_4$ gravity
background from IIA string theory. The D4 branes background of QCD$_4$ with Euclidean signature have two compactified
directions and D8 branes are flavors~\cite{S-S:0412}. There are two main phases, confinement and deconfinement phases
in this holographic model. In the present paper, we will investigate the properties of a probe baryon in these two
phases, which is an extended work on baryon probe, of the former work~\cite{Peeters:2006}. 
 It's believed that heavy-quark bound state can survive in a quark gluon plasma at a
temperature which is higher than the confinement/deconfinement transition temperature~\cite{de Forcrand:2000jx}. We
want to have a closer look at the baryon configuration in these different phases, especially in the deconfined phase of
a hot quark-gluon plasma.

The present paper is organized as follows. In section 2, we review different phases of QCD$_4$ at different
temperatures. We also review a point brane + strings model and a DBI brane model of a holographic baryon in phases in
QCD$_4$ background and main properties of AdS$_6\times$S$^4$ background we will use. In section 3, we propose a new
baryon configuration in different backgrounds. In section 4, we study the screening length, baryon mass and interacting
energy in QCD$_4$ background. In the last section, we study the baryon properties and define a new screening length in
warped AdS$_6\times$S$^4$ background.

\section{Review of different background phases and baryon probe model}
\subsection{Different phases of D4 branes background}
\paragraph {QCD$_4$ background}

We summarize different geometry backgrounds from the D4-D8 branes model~\cite{S-S:0412}.
The main success of this model is the prediction of the spectrum of
low energy hadrons and their dynamics, which are non-perturbative
properties of QCD. It's a good choice for the top-down holographic
QCD model since it contains chiral fermions and an apparent chiral
symmetry breaking mechanism. Gluons are represented in terms of
fluctuating modes of open strings on $N_c$ D4 branes. Massless
quarks are represented in terms of fluctuations modes of open
strings connecting $N_c$ D4 branes and $N_f$ D8($\overline{D8}$)
branes.\footnote{We do not consider these light quarks in the same
way of the work ~\cite{Peeters:2006} and just consider a probe
baryon composed with heavy quarks in our present work.} The
supergravity description of $N_c$ D4 branes and gauge field
description of the D4 branes give the gauge/gravity duality. If
$N_c>>N_f$, D8 branes which give the flavor freedom have no
backreaction to the background geometry and can be considered as
probes. When the excitations on D4 branes are very heavy, there will
be a black hole in the bulk, which corresponds to the thermal gauge
field on the boundary. At different temperatures, there appear
different gravity backgrounds~\cite{Aharony:2006da}.
The confining background metric of Euclidean model at zero temperature
 is
 \be\label{confined background}
 ds^2=({r\ov R})^{3/2}(dt_E^2+d\vec x^2+f(r)dx_4^2)+({R\ov r})^{3/2}
 ({dr^2\ov f(r)}+r^2d\Omega_4^2)
 \ee
 with a dilaton and a 4-form RR field strength
 \bea
 \begin{split}
 e^{\phi}=g_s({r\ov R})^{3/4},&\quad\quad F_4={2\pi N_c\ov
 V_4}\epsilon_4.\\
 \end{split}
 \eea
 where $f(r)=1-({r_{c}\ov r})^3$. The duality relation is $R^3=\pi g_sN_cl_s^3$. $g_s,N_c,l_s$ are string coupling constant,color number and string
 length scale, respectively. $\vec x=x_{1,2,3}$, $r$ is the radial coordinate and
 $\Omega_4$ is four angle coordinates, in the $x_{5,6,7,8,9}$ space.
 $V_4=8\pi^2/3$ is the volume of the unit S$^4$ and $\epsilon_4$ is
 the volume 4-form. To cancel the conical singularity at $r=r_{c}$,
 the period of $\delta x_4$ in the compactified direction must be
 \be
 \delta x_4={4\pi \ov 3}({R^3\ov r_{c}})^{1/2}.
 \ee
 The Kaluza-Klein mass is defined as
 \be
 M_{KK}\equiv{2\pi\ov \delta x_4}={3 \ov 2}({r_c\ov R^3})^{1/2}.
 \ee
 In this phase, we can obtain the glueball and meson spectra by
 computing the fluctuation of background supergravity fields
 and fields on the flavor branes respectively. Their spectra are discrete, which
 shows that the system is confined.
 We regard the Hawking temperature of the background
 as the temperature of thermal field. We have $T=1/\beta$, where
 $\beta$ is the period of Euclidean time.
 At high temperature region, the phase is deconfined. The
 background geometry contains a black hole:
 \bea
ds^2=({r\ov R})^{3/2}(f(r)dt_E^2+d\vec x^2+&dx_4^2)+({R\ov r})^{3/2}
 ({dr^2\ov f(r)}+r^2d\Omega_4^2)
 \eea
 with a dilaton and a 4-form RR field strength
 \bea
  e^{\phi}=g_s({r\ov R})^{3/4},&\quad\quad F_4={2\pi N_c\ov
 V_4}\epsilon_4.
 \eea
 where $R^3=\pi g_sN_cl_s^3,\quad f(r)=1-({r_0\ov r})^3$.
 The $t_E$ must have period
 \be
 \delta t_E={4\pi\ov 3}({R^3\ov r_0})^{1/2}={1\ov T}.
 \ee
 This background metric has the same form as the confining one (\ref{confined background}),
 only with the exchanging of $t_E$ and $x_4$.

\paragraph{Warped AdS$_6\times$ S$^4$ background}

Massive type IIA supergravity admits a warped AdS$_6$$\times$ S$^4$ vacuum solution, which is expected to be dual to an
$\mathcal {N}$=2, D=5 super-conformal Yang-Mills theory. This background is supported by Ramond-Ramond field strengths,
in addition with a non-constant dilaton. The AdS$_6\times$ S$^4$ metric is warped, with a warped factor which becomes
singular on the equator of S$^4$. Thus the geometry really corresponds to a hemisphere instead of the full S$^4$. This
background arises as the near-horizon geometry of a semi-localized D4-D8 system~\cite{Brandhuber:1999np,Youm:1999zs}.
In the string frame, this solution is given by~\cite{Cvetic:1999un,Chong:2004kf}
 \be\label{Lumetric}
 ds_{10}^2=(\cos\theta)^{-1/3}[ds_{AdS_6}^2+2g^{-2}(d\theta+\sin^2\theta
 d\Omega_3^2)]\;.
\ee Four form field strength and the dilaton take the forms \be
 F_{(4)}={5\sqrt{2}\ov 6}g^{-3}(\cos\theta)^{1/3}\sin^3\theta
 d\theta \wedge \Omega_{(3)}\;,\quad e^\phi=(\cos\theta)^{-5/6},
 \ee where $\Omega_{(3)}$ is three sphere volume form.
 The AdS black hole metric is
\bea
 ds_{AdS_6}^2=-f(r)dt^2+{dr^2 \ov f(r)}+r^2(\Sigma_{i=1}^{4}dx^idx^i)\;.
 \eea where $f(r)=g^2r^2-{\mu \ov r^3}$, $g$ and $\mu$ are two parameters independent on $r$.

\subsection{Point brane + hanging strings configuration for baryon}
Holographic baryons have configurations of hanging strings attached to a compact vertex brane. While baryons in the
boundary field are composed of external heavy quarks. In probe limit, each component quark attached with a fundamental
string is considered as a probe string in the bulk. Since these fundamental strings connect with a vertex brane, we can
see this vertex brane as a probe. In recent works~\cite{Liu:2008bs,Li:2008py},
 the configuration of open strings+compact D5 brane wrapped on the
S$^5$ were analyzed. The background is given by
 \bea
 ds^2=-f(r)dt^2 +\frac{r^2}{R^2}dx_3^2 +\frac{r^2}{R^2}\left(
d\rho^2+\rho ^2d\theta^2\right)+\frac{1}{f(r)}dr^2+R^2 d\Omega^2_5\;.
  \eea where $ f(r)=\frac{r^2}{R^2}\left(1-\frac{r_0^4}{r^4}\right)$.
 The action of baryon is summation of $N_c$ fundamental strings and D5 brane
 \bea  S_{total} =
  \sum_{i=1}^{N_c}  &S_{string}^{(i)} + S_{D5} \;.
  \eea
  where the action of D5 is given by a massive point action in gravity field
 \bea S_{D5} =\frac {\VV (r_e) \TT V_5 }{(2 \pi)^5 \alpha'^3 }\;.
 \eea
The screening length of baryon with finite moving speed in a plasma
was computed~\cite{Liu:2008bs}. High spin baryons in the
AdS$_5\times S^5$ were investigated and the angular velocity
dependence of screening length and J-E$^2$ behavior were computed
numerically~\cite{Li:2008py}. However, all these computations were
done in the conformal and supersymmetric background, and the D5
brane is treated as a massive point in AdS$_5$. The QCD$_4$
background is a good choice for the non-supersymmetric and
non-conformal background. We can construct a new configuration with
a DBI D4 brane and N$_c$ strings in this background.

\subsection{A brane model for baryon vertex}
Ten years ago, people believed that main information of a holographic baryon was hidden in the vertex brane. We can
indeed get interesting information of a holographic baryon from the DBI+CS action of a compact brane. We take a
confining background for example as follow:
 \be
 ds^2=({r\ov R})^{3/2}(dt_E^2+d\vec x^2+f(r)dx_4^2)+({R\ov r})^{3/2}
 ({dr^2\ov f(r)}+r^2d\Omega_4^2)
 \ee
where $r$ is the radial coordinate and the embedding function of the
compact D4 brane wrapped on the S$^4$ can only be determined by
$r(\theta)$. And the gauge field on the single D4 brane can also be
written as $A(\theta,t)$ for symmetry. The action of the D4 brane
obtained from the induced metric is given
by~\cite{CGST:9810,CGST:9902}\be
 S_{D4}=-T_4\int d^5\xi e^{-\tilde{\phi}}\sqrt{-\det(g+F)}+T_4\int A_{(1)}\wedge G_{(4)}
\ee From this action, we can solve the equation of motion for gauge
field and find the embedding function $r(\theta)$.

It is argued that a closed curve $r(\theta)$ corresponds to a real
baryon vertex. The vertex brane dynamics can be also solved in
 a deconfined phase, but there is no closed
 solution~\cite{CGST:9810}. However,
 there is abundant evidence to support that
heavy-quark bound states can survive in a quark-gluon
plasma~\cite{de Forcrand:2000jx}, which is in the deconfined phase
from the lattice results. So there is an apparent paradox. To solve
this problem, we try the deconfined warped AdS black hole background
in subsection 3.3 in the present paper, where the warped AdS example
reflects more faithfully the physics expected in actual QCD.

\section{DBI brane + N$_c$ strings configuration in different phases}
In this section, we propose a DBI brane + N$_c$ strings
configuration for a holographic baryon probe and investigate its
properties in different phases of D4 branes background.
\subsection{DBI brane + N$_c$ strings in confined QCD$_4$ background}
We start from the confined background: \bea
 ds^2=({r\ov R})^{3/2}(dt_E^2+d\vec x^2+f(r)dx_4^2)+({R\ov r})^{3/2}
 ({dr^2\ov f(r)}+r^2d\Omega_4^2)\;.
 \eea
This metric with periodic Euclidean time and a compactified space
direction $x_4$ corresponds to four dimensional boundary thermal
field. We consider a static baryon in the thermal field as hanging
open strings attached to a D4 brane wrapped on the $S^4$. Open
strings connect flavor branes and the single compact D4 brane. In
this configuration, we denote the world volume coordinates of D4
brane as $(t, \theta, \alpha, \beta, \gamma)$. And the embedding
function is $r=r(t, \theta, \alpha, \beta, \gamma)$ and the U(1)
gauge field on the D4 brane is $A_\mu=A_\mu(t, \theta, \alpha,
\beta, \gamma)$\;. The induced metric on D4 brane is \be
 ds_{D4}^2=-({r\ov R})^{3/2}dt^2+({R\ov r})^{3/2}[({r'{}^2\ov f(r)}+r^2)d\theta^2+r^2\sin^2\theta d\Omega_3^2]
\ee The action of the compact D4 brane can be given as \be S_{D4}=-T_4\int d^5\xi
e^{-\widetilde{\phi}}\sqrt{-\det(g+F)}+T_4\int A_{(1)}\wedge G_{(4)}\;, \ee where, $g_{\mu\nu}$ is the induced metric
on the D4 brane. $F=dA$, $T_4=1/(g_s(2\pi)^4l_s^5)$ is D4 brane tension, and the last term is Wess-Zumino term. We
suppose that the D4 brane wrapped on the S$^4$ has a SO(4) symmetry, and the embedding function and gauge field depend
only on $(t, \theta)$\;.  For a static configuration, we have $r=r(\theta)$ and $A_t=A_t(\theta)$\;. We can rewrite the
action by performing Legendre transformation~\cite{CGST:9902}:\footnote{Since this is a static solution, we ignore the
time component.} \be\label{EL}
 \mathcal {H}  =T_4\Omega_3R^3\int
 d\theta\sqrt{r^2+f(r)^{-1}r'(\theta)^2}\sqrt{D(\theta)^2+\sin^6\theta}\;,
\ee where the displacement $D$ satisfies \be\label{D}
 \p_\theta D=-3\sin^3\theta\;,
\ee which comes from the equation of motion of the gauge field.
Solving equation(\ref{D}), we get \be
 D(\theta)=3\cos\theta-\cos^3\theta-2+4\nu\;.
\ee $\nu$ in the constant of integration is a parameter $0\leq\nu\leq1$, which controls the number of Born-Infeld
strings emerging from the D4-brane at each pole of the S$^4$ ($\theta=\pi $ and
$\theta=0$)~\cite{CGST:9902}~\cite{CGST:9810}. We find that the spikes at $\theta=0$ and $\theta=\pi$ have the same
asymptotic `tension' as $\nu N_c$ and $(1-\nu)N_c$ fundamental strings, respectively. We hope that the solution which
contains a vertex brane with excited string-like spikes can be obtained from the total action of  vertex brane. But
there remain lots of challenges~\cite{CGST:9810,CGST:9902,GRST:9907,Ima:0410,Kim:0002}. So we always use FBC to cancel
the singularity of vertex brane.

\begin{figure}[t]
\centering
  \includegraphics*[width=0.6\columnwidth]{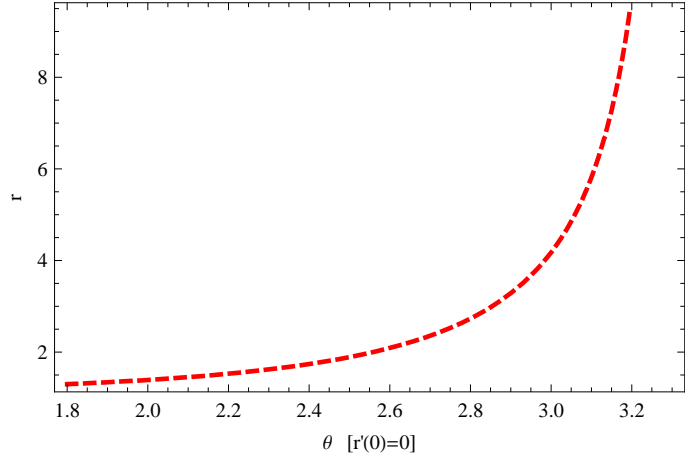}
  \caption{\small Embedding function of D4 brane in $r-\theta$ plane    }\label{r_con}
\end{figure}

Now we give the numerical result in Figure \ref{r_con}. We set $r'(0)=0$ and consider $\nu=0$, which means that $N_c$
fundamental strings all connect with the north pole of S$^4$.
The $N_c$ fundamental strings hang from flavor brane (we just
consider single flavor). By the following string embedding \be
\tau=t,\quad\sigma=r,\quad\rho=\sqrt{x_1^2+x_2^2}=\rho(r),\quad
x_3=\textrm{constant}\;,\ee where $x_i(i=1,2,3)$ is the boundary
spatial direction, the action can be written as \be
 S_{string}=\frac {1}{2 \pi \alpha' }\int dt \int_{r_e}^{r_\Lambda}
 dr
\sqrt{-\det[h_{ab}]} \;, \ee thus the string world sheet Lagrangian density is\be
 \LL=\sqrt{-\det[h_{ab}]}\;,
\ee
 where $h_{ab}=g_{\mu\nu}{\partial x^\mu
\partial x^\nu \ov
\partial \sigma^a \partial \sigma^b}$.
The total action is \be S_{total} =
  \sum_{i=1}^{N_c}  S_{string}^{(i)} + S_{D4}\;.
\ee To obtain the force balance condition (FBC), we rewrite the
action of the D4 brane: \be
 S'_{D4}=\int dt \mathcal {H}
\ee Extremizing $S_{total}$ with respect to $r_e$, we get FBC
\be\label{FBC}
  \sum_{i=1}^{N_c} H^{(i)} \biggr|_{r_e} = \Sigma \;,
 \ee
 where
 \be\label{H}
 H^{(i)} \equiv
 \LL^{(i)} -  \rho'^{(i)} {\p \LL^{(i)} \ov \p \rho'^{(i)}} \;,
 \ee
 \be\label{sigma}
 \Sigma \equiv  {2\pi \alpha'\ov \TT}{\p S'_{D4}\ov \p r_e}=2\pi \alpha'{\p \mathcal {H} \ov \p
   r_e} \;.
\ee For a given $r_0$ we can get the value of $r_e=r(\pi)$ from the
EOM of $r(\theta)$. Then we can get the embedding function of $N_c$
strings with FBC.

 We consider a D8 brane as a flavor brane. The interaction between
the D8 brane and the background $N_c$ D4 branes makes the D8 brane
embed nontrivially in the confined geometry. The DBI action of the
D8 brane is \be
 S_{D8}=T_8\int d\xi^9\sqrt{-\det(g+F)}\;,
\ee where the D8 brane extends ($t,\vec x, x_4, \Omega_4$). The gauge field fluctuations on the flavor brane always
correspond to vector mesons in the boundary field. In the present paper, we ignore these fluctuations. After the ansatz
of the D8 brane, the embedding function is only determined by $x_4(r)$ and we can obtain the solution by solving the
equation of motion with an initial condition.
In the D4-D8 system, the heavy quarks have no backreaction
to the system, so we ignore the pull force of hanging strings to the flavor brane. As discussed in our former
work~\cite{Li:2008py}, we can obtain solutions and define screening length.

Let us turn to the solution of the compact D4 brane we obtained in
Figure $\ref{r_con}$. We will get a very large $r_e$ if we change
the initial $r(0)$ to a suitable one. The cusp of the D4 brane will
replace the fundamental strings and touch the flavor brane if
$r_e\geq r_\Lambda$. But, since we assume that $N_c$ probe quarks (
not dynamic ) have no backreaction to D8, here $N_c$ hanging strings
do not need to be replaced by the deformed D8 brane. It's very
different from the case of finite quarks density where the deformed
flavor brane always replaces many open strings connected with the
horizon. In experiments, component quarks in probe baryons which can
survive in the QGP are much heavier than the background quarks or
gluons in quark gluon plasma. There will be some apparent properties
of the probe in the medium, especially when we boost or rotate it.
These properties are important signals for hot strongly coupled
plasma.

\subsection{DBI brane + N$_c$ strings in deconfined QCD$_4$ background}

 As the temperature rises up, the D4-D8 system will undergo a first
order phase transition, where the gluonic degree of freedom get
deconfined. The corresponding background geometry becomes:
 \be
 ds^2=({r\ov R})^{3/2}(f(r)dt_E^2+d\vec x^2+dx_4^2)+({R\ov r})^{3/2}
({dr^2\ov f(r)}+r^2d\Omega_4^2)
 \ee
 The investigation of baryon probes in this phase is similar to the
 investigation in confined phase. The induced metric on the
vertex D4 brane is
 \be
ds_{D4}^2=-({r\ov R})^{3/2}f(r)dt^2+({R\ov r})^{3/2}[({r'^2\ov
f(r)}+r^2)d\theta^2+r^2\sin^2\theta d\Omega_3^2]
 \ee
 The DBI action of the compact D4 brane is
 \be
S_{D4}=-T_4\int d^5\xi e^{-\tilde{\phi}}\sqrt{-\det(g+F)}+T_4\int
A_{(1)}\wedge G_{(4)}\;,
 \ee
Thus using the same method in the confined case, we obtain the energy function\be
 \mathcal {H}=T_4\Omega_3R^3\int
 d\theta\sqrt{f(r) r^2+r'{}^2}\sqrt{D(\theta)^2+\sin^6\theta}\;,
\ee

Among solutions of $r(\theta)$ from the above Lagrangian, we can not
find a closed one. $r(\pi)$ of all solutions run to
infinity.\footnote{We do not consider this kind of unclosed
solutions here due to the following reasons. One is that for not
very heavy probe quarks these solutions are usually not regarded as
baryons since quarks and gluons are deconfined in this phase.
Another reason is that we don't know how to analyze properties of
baryon probes in detail by these solutions, though we can also think
that for very heavy probe quarks, this D4-brane ``spike'' by itself
represents a bundle of N strings and these solutions can be regarded
as baryons within which the quarks have coalesced and are no longer
individually discernible ~\cite{CGST:9902}.} But we can still try to
get the baryon configuration as in ~\cite{CGST:9902} where the
background has no $x_4$ direction and becomes Euclidean effectively.
And an effective baryon there there has a transformed Lagrangian
which is similar to (\ref{EL}).

\subsection{DBI brane + N$_c$ strings in warped $AdS_6\times S^4$}

\begin{figure}[t]
\centering
  \includegraphics*[width=0.6\columnwidth]{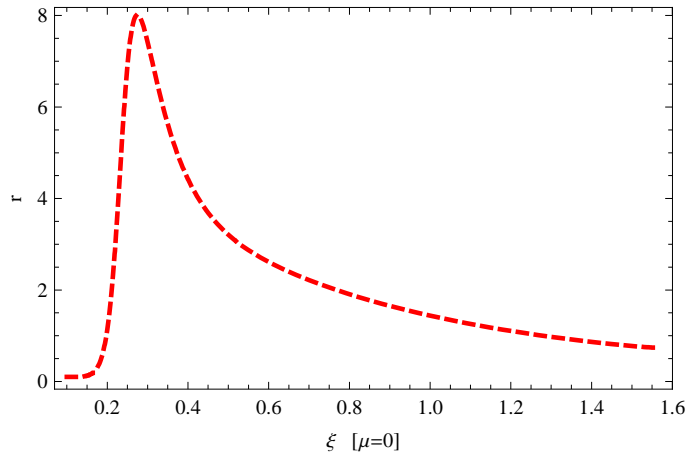}
  \caption{\small Embedding function of D4 brane in $r-\xi^1$ plane}\label{r_xi}
\end{figure}

Massive type IIA supergravity admits a warped AdS$_6$$\times$ S$^4$ vacuum solution, which is expected to be dual to an
$\mathcal {N}$=2, D=5 super-conformal Yang-Mills theory. We study a DBI brane + N$_c$ strings configuration for baryon
in this background. In Eq.(\ref{Lumetric}), we note that the metric is singular at $\theta={\pi\ov 2}$, thus $\theta$
covers $[0, {\pi\ov 2})$ in a hemisphere instead of the full S$^4$. We shall study DBI vertex brane + hanging strings
configuration in this geometry, which corresponds to a baryon state in $\mathcal{N}$=2, D=5 super-conformal Yang-Mills
theory. We denote coordinates of three sphere as $(\alpha, \beta,\gamma)$ and world volume coordinates of a vertex D4
brane as $(\tau,\xi^1,\xi^2,\xi^3,\xi^4)$. By the following consistent ansatz that describes the embedding D4 brane \be
\tau=t,\quad \xi^1=\theta,\quad \xi^2=\alpha,\quad \xi^3=\beta,\quad \xi^4=\gamma, \quad r=r(\xi^1)\ee we write the
induced metric of D4 brane \be ds_{D4}=(\cos\xi^1)^{-1/3}[-f(r)d\tau^2+({r'^2(\xi^1)\ov
f(r)}+2g^{-2})d^2\xi^1+2g^{-2}\sin^2\xi^1 d\Omega_3^2]\;. \ee The world volume action of D4 brane contains a DBI term
and a Chern-Simons term \be
 S_{D4}=S_{DBI}+S_{CS}\;,
\ee where
\bea \begin{split}
 S_{DBI}=-T_4\int d\tau d^4\xi&\sqrt{-\det(g+F)} \\
=-T_4g_s\Omega_3({\sqrt{2}\ov g})^3\int &d\tau
d\xi^1\sin^3(\xi^1)\\
&\times\sqrt{r'(\xi^1)+2g^{-2}f(r)-\cos(\xi^1)^{2/3}F_{t\xi^1}^2}\;.
\end{split} \eea
The Chern-Simons coupling term is \bea
\begin{split}
S_{CS}&=T_4\int A \wedge\mathscr{P}(F_4)\\
&=-{5\sqrt{2}\ov 6}g^{-3}\Omega_3\int d\tau d\xi^1 A_t\cos(\xi^1)^{1/3}\sin^3(\xi^1)\;,
\end{split} \eea
where $\Omega_3$ is the volume of unit $S^3$ . We obtain the vertex D4 brane Lagrangian density along $\xi^1$ \bea
\begin{split}
 \mathcal
 {L}_{D4}=\sin^3(\xi^1)\cos(\xi^1)^{1/3}\sqrt{[r'(\xi^1)+2g^{-2}f(r)]\cos(\xi^1)^{-2/3}-F_{t\xi^1}^2}+{5\ov
 12g_s}A_t\;. \end{split}
\eea Thus the action of vertex D4 brane can be written as\be
 S=-T_4\Omega_3g_s({\sqrt{2}\ov g})^3 \int d\tau\mathcal
 {L}_{D4}\;.
 \ee
 The equation of motion of $A_t$ is given by

\be
{\p  \mathcal
 {L}_{D4}\ov \p F_{t\xi^1}}=-D(\xi^1)\;,
 \ee
 where
 \be
 \p_{\xi^1}D(\xi^1)={5\ov
 12g_s}\sin^3(\xi^1)\cos(\xi^1)^{1/3}\;.
 \ee
To solve $D(\xi^1)$, we denote $y=\sin(\xi^1)$, then $D(\xi^1)$ is
given by \be D(y)={5\ov
 12g_s}{1\ov 1729}3y[-(1-y^2)^{2/3}(135+105y^2+91y^4)+135F[{1\ov 2},{1\ov 3},{3\ov
2},y^2]]+C \ee The Legendre  transformation of the Lagrangian can help to eliminate the gauge field. We obtain an
energy function of the embedding coordinate $r(\xi^1)$ only: \be \mathcal {H}=-T_4\Omega_3g_s({\sqrt{2}\ov g})^3\int
d\xi^1\sqrt{[r'(\xi^1)+2g^{-2}f(r)]\cos(\xi^1)^{-2/3}[\sin^6(\xi^1)\cos(\xi^1)^{2/3}+D^2]} \ee From this Lagrangian, we
get the EOM of $r(\xi^1)$

\be {\p \mathcal {H}\ov \p r}-\p_{\xi^1}{\p \mathcal {H}\ov \p r'}=0 \ee To see the result quickly, we solve the EOM
numerically in Figure \ref{r_xi}. From Figure \ref{r_xi}, we find that for $\xi^1\in[0,{\pi\ov 2})$, $r$ has finite
values. Since the hemisphere geometry corresponds to the background with $\theta\in[0,{\pi\ov 2})$, $\xi^1$ can only
take values between $0$ and ${\pi\ov 2}$. The warped AdS black hole corresponds to a deconfined gauge field theory, so
we obtain closed vertex D4 brane solutions from the DBI+CS action naturally.

\section{Baryon properties in QCD$_4$ background}


Baryon probes in quark gluon plasma were first investigated in works~\cite{Alberto:0611,Liu:2008bs,Li:2008py}, where
the background is AdS$_5$$\times$S$^5$ dual to the $\mathcal {N}=4$ supersymmetric Yang-Mills gauge field, and the
vertex D5 brane sits a point in the AdS space in~\cite{Liu:2008bs,Li:2008py}. In D4-D8 system, we consider that the
trajectory of the vertex D4 brane is not a trivial point and the embedding function depends on $r(\theta)$. We are
interested in the properties of baryon probe of this configuration. We analyze the screening length and baryon mass
behavior in this section.
\subsection{Screening length and baryon melting} In the former work~\cite{Liu:2008bs,Li:2008py}, screening length of
baryon was defined as the largest value of the boundary quark separation when we change the position of the bottom of
hanging strings $r_e$. From our solutions, we find the trajectory of vertex brane in the bulk radial direction is a
finite line. We defined the position of cusp of vertex brane as the furthest position which can be probed by the
baryon. Thus, we obtain the $r_e$ dependence of the quark separation $l_q$ and get the value of screening length $l_s$.

We consider boosted and rotating quarks and use the following background:\footnote{Corresponding hanging strings rotate
rigidly in the bulk, which is like the meson case in work~\cite{Peeters:2006} and different from the strings with
rotating ends in work~\cite{Fadafan:2008bq}.}
\begin{equation}
\label{boostmetric} ds^2=(\frac{r}{R})^{3/2}(-dt^2 +dx_3^2)
+(\frac{r}{R})^{3/2}\left( d\rho^2+\rho ^2d\theta^2\right)+({R \ov
r})^{3/2}\frac{dr^2}{f(r)}
\end{equation}

The metric is invariant when it is boosted in $x_3$ direction. So
properties of a baryon are independent on wind in $x_3$ direction.
We now focus on baryon configurations rotating in the $x_1-x_2$
plane at angular velocity $\omega$.
For symmetry, each string can be described by $\rho(r)$. By the following consistent ansatz \be \label{parameters}\tau
= t \;, \qquad \sigma=r\;, \qquad \theta =\omega t\;, \qquad \rho=\rho(r)\;, \ee We can write the action of a single
string is: \be
 S_{string}={1\ov 2\pi\alpha'}\int
 dt\int_{r_e}^{r_\Lambda}dr\sqrt{-\det[h_{ab}]}
\ee $h_{ab}$ can be read from the induced metric
 \be
ds^2=({r\ov R})^{3/2}(\rho^2\omega^2-1)dt^2 +\biggr[({r\ov R})^{3/2}\rho'^2+({R\ov r})^{3/2}{1\ov f}\biggr]dr^2\;. \ee
The string world sheet Lagrangian density \be\mathcal {L}_{string}=({r\ov
R})^{3/2}\sqrt{(1-\rho^2\omega^2)(\rho'^2+{R^3\ov r^3-r_{c}^3})}\;.
 \ee
 To solve the equation of motion
\begin{equation}\label{EOM}
\biggr[{\p\ov\p\rho(r)}-{\p \ov \p r}{\p\ov\p\rho'(r)}\biggr]\LL=0
\;,
\end{equation}
We need two boundary conditions $\rho(r_e)$ and $\rho'(r_e)$. Where the constraint at $r_e$ can be obtained from the
FBC in eqa (3.11):
 \be
 \LL -  \rho' {\p \LL \ov \p \rho'}\biggr|_{r_e}={2\pi\alpha'\ov N_c}
 {\p \mathcal {H} \ov \p r_e} \;,
 \ee
 The FBC turns out to be
 \be
({r\ov R})^{3/2}\frac {({R^3\ov
 r^3-r_{c}^3})}{\sqrt{(\rho'^2+{R^3\ov
 r^3-r_{c}^3})}}\biggr|_{r_e}={8\pi\alpha'T_4\Omega_3R^3\ov
 N_c}{f^{-1}r'\ov
 \sqrt{r^2+f^{-1}r'{}^2}}\biggr|_{r=r(\theta=\pi)}\;.
 \ee
 Finally, by analyzing the solutions we define the screening length as the critical value of
 the boundary quark separation and find the $\omega$ dependence of screening length $l_s$.
 The screening length $l_s$ shows that if heavy quarks have
 enough kinetic energy, they may break away from each other. It
 means that baryons dissociate.

\begin{figure}[t]
\centering
  \includegraphics*[width=0.6\columnwidth]{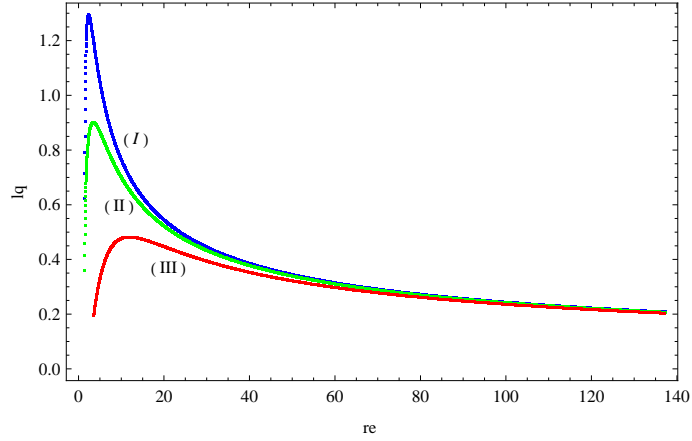}
  \caption{\small $r_e$ dependence of $l_q$ at different values of $\omega$. Curves I, II, III correspond to $\omega=0,0.5,1$ respectively. Values of $l_q$ and $r_e$ are determined assuming $r_0$ as unit.}\label{lq_re}
\end{figure}

\begin{figure}[t] \centering
  \includegraphics*[width=0.6\columnwidth]{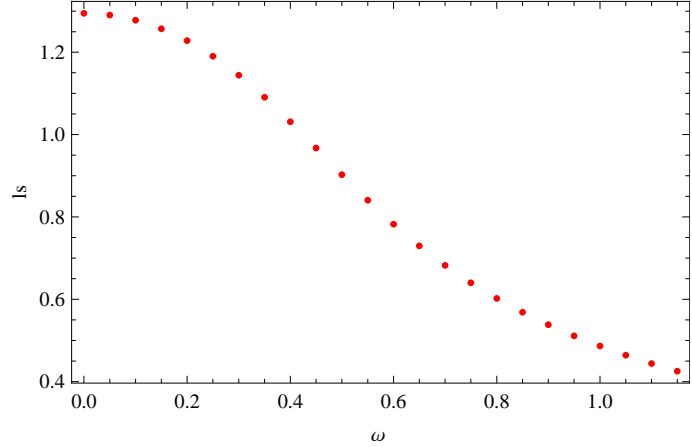}
  \caption{\small $\omega$ dependence of $l_s$.}\label{ls_omega}
\end{figure}

 To obtain the screening length of the DBI brane + strings
 configuration, we calculate the $r_e$ dependence of $l_q$
 numerically which has been discussed more carefully in ~\cite{Li:2008py}. The
 results with different $\omega$ are shown in Figure \ref{lq_re}.
We choose the maximal value of $l_q$ (also the critical value) as
the screening length for baryons. And we think that only the right
part of each curve beside the highest point contains the points
which correspond to real baryon configurations. The $\omega$
dependence of $l_s$ can be given in Figure \ref{ls_omega}.

\subsection{Baryon mass and interaction potential} 
\begin{figure}[t]
\centering
  \includegraphics*[width=0.6\columnwidth]{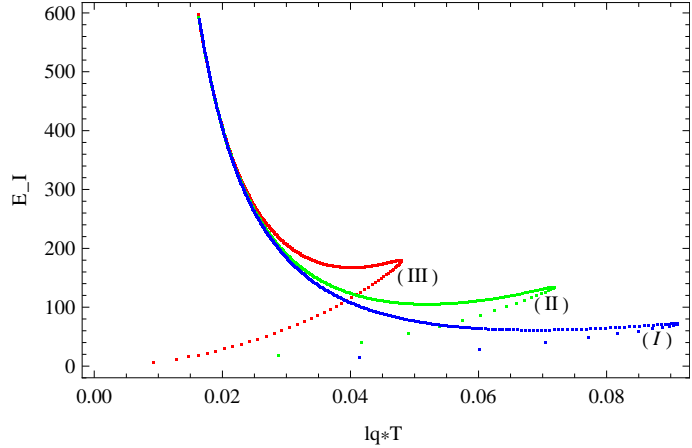}
  \caption{\small $l_q$ dependence of interaction potential of $N_c$ quarks. Curves I, II, III correspond to $\omega=0.3,0.5,0.8$ respectively. For simplicity, the value of $E_I$ in the figure is actually $E_I*(2\pi\alpha'/N_c)$ in our paper.}\label{EI_lq}
\end{figure}

Now we want to give the
definition of baryon mass and interaction potential. In a very general way, baryon mass is given by summation of the
energy of $N_c$ strings and the vertex brane.  \be\label{Etotal}
 E_{total}=N_cE_{string}+E_{D4},
\ee where \be\label{stringenergy} E_{string}=\omega {\p L \ov
\p\omega}-L\;, \quad E_{D4}=\mathcal {H}\;,
 \ee
 The string Lagrangian is
 \be
L=\frac {1}{2\pi \alpha'}\int_{r_e}^{r_{\Lambda}}dr ({r\ov
R})^{3/2}\sqrt{(1-\rho^2\omega^2)(\rho'^2+{R^3\ov
 r^3-r_{c}^3})}\;.
 \ee
 In order to obtain the interaction potential, we analyze the free quarks
 case, in which $N_c$ strings hang from the boundary to $r_{c}$ and
 compact D4 brane almost wrapped on the $r=r_0$.
 The so called interaction potential is given by subtracting the energy of the free strings
 and the corresponding vertex brane. Assuming the radial position of D8 is $r_{\Lambda}$, the radial distance
 of $r_{c}$ and D8 is $r_{\Lambda}-r_{c}$,  and the energy of single quark is given
 by \footnote{Actually, there is no free quark in this confined phase, to calculate the interaction energy in the string picture,
  we should subtract mass of $N_c$ quarks, which are not real physical objects in this confined phase but a reference.}
 \be
  E_q={1\ov 2\pi\alpha'}\int_{r_{c}}^{r_\Lambda}dr\;,
 \ee
 and the energy of
 initial brane which is very close to the ``cut off position''($r=r_{c}$) is given by\be
  \mathcal {H}_0=T_4\Omega_3R^3r_0\int
 d\theta\sqrt{D(\theta)^2+\sin^6\theta}\;.
 \ee
 Then the interaction potential of baryon is\footnote{We should note that interaction potential here is different from
  binding energy, the rigid rotating effect of strings  and the vertex brane contribute more energy to make $E_I$ positive.
   }

 \be
  E_I=E_{total}-N_cE_q-\mathcal {H}_0\;. \ee
 We give the numerical results about $l_q$ dependence of $E_I$ at
 different values of $\omega$ in Figure \ref{EI_lq}.
The Figure \ref{EI_lq} shows that we should choose the low energy
branch for a given $l_q$. Comparing with Figure \ref{lq_re}, we find
that the low energy branch in Figure \ref{EI_lq} corresponds to the
left branch in Figure \ref{lq_re}. It implies that we should choose
the smaller $r_e$ for a given $l_q$. The main reason is that larger
$r_e$ corresponds to larger D4 brane energy. Actually, for a given
$r_0$ larger than a critical value, the D4 brane will be enlonged to
the boundary D8 brane and the length of hanging strings becomes
zero. Thus we can not obtain a closed brane configuration, which is
similar to the deconfined case. When the background becomes closer
to a deconfined one, quarks will try to run away from each other. In
our confined background, it corresponds to the vertex  D4 brane
enlonged to attach to the boundary brane.
\subsection{Discussion} After a
simple discussion of point brane + $N_c$ strings in AdS$_5\times$S$^{5}$ in a former work~\cite{Li:2008py}, we extend
the investigation to a new configuration of a DBI brane + $N_c$ strings in QCD$_4$ background. Our investigation
implies that some properties such as screening length still exists in the new configuration. But the baryon mass and
interaction potential are very different from the simple model with a point vertex brane in the AdS bulk. In the
QCD$_4$ background, there is no horizon and we find that the minimal energy state corresponds to the point-like D4
brane in the AdS gravity background, connected to $N_c$ hanging strings with the largest length. It appears very
different from the $N_c$ strings + point D5 brane as before~\cite{Li:2008py,Liu:2008bs}, where the energy of the D5
brane depends only on $r_e$. The energy of D4 brane dominates the interaction potential in this case.

\section{ Baryon properties in warped AdS$_6\times$S$^4$ background}
After analyzing the properties of holographic baryon probe in QCD$_4$ background, we want to see properties of our
solution in warped AdS$_6\times$S$^4$. We consider boosted and rotating quarks and use the following background: \be
ds^2=(\cos\theta)^{-1/3}[-(g^2r^2-{\mu \ov r^3})dt^2+r^2dx_3^2+{dr^2 \ov (g^2r^2-{\mu \ov
r^3})}+r^2(d\rho^2+\rho^2d\theta^2)] \ee If we stand in  the rest frame of quarks, we will feel a moving plasma wind.
We now focus on baryon configurations rotating in the $x_1-x_2$ plane with angular velocity $\omega$ in the plasma
moving with a wind velocity $v=- \tanh\eta\;$ in the $x_3$ direction.
For symmetry, each of strings can be described by $\rho(r)$. By
the following consistent ansatz \be \label{parameters}\tau = t \;,
\qquad \sigma=r\;, \qquad \theta =\omega t\;, \qquad
\rho=\rho(r)\;,\quad x_3=x_3(\sigma)\;, \ee We can write the action
of a single string is: \be
 S_{string}={1\ov 2\pi\alpha'}\int
 dt\int_{r_e}^{r_\Lambda}dr\sqrt{-\det[h_{ab}]}={1\ov 2\pi\alpha'}\int
 dt\int_{r_e}^{r_\Lambda}dr\mathcal {L}_{string}\;,
\ee $h_{ab}$ can be read from the induced metric on the string world
sheet(more details can be read from appendix). All hanging strings
attach to the highest point of vertex solution in $r$ direction,
then the FBC turns to be \be
 L -  \rho' {\p L \ov \p \rho'}\biggr|_{r_e}={2\pi\alpha'(\cos\theta)^{1/3}\ov N_c}
 {\p \mathcal {H} \ov \p r_e} \;,
\ee where $r_e=r_{top}$(with $r'(\xi^1)=0$) in the brane solution in
Figure \ref{r_xi}. To calculate the ${\p \mathcal {H} \ov \p r_e}$,
we rewrite the energy function of D4 brane as \be
\begin{split}\mathcal {H}
=-T_4\Omega_3g_s({\sqrt{2}\ov g})^3\int
dr\sqrt{[1+2g^{-2}f\xi'^1(r)]\cos(\xi^1)^{-2/3}[\sin^6(\xi^1)\cos(\xi^1)^{2/3}+D^2]}\;.\footnote{$\xi'^1(r)={\p\xi^1\ov
\p r.}$}
\end{split}
\ee We denote \be \mathscr{L}=
\sqrt{[1+2g^{-2}f(r)\xi'^1(r)]\cos(\xi^1)^{-2/3}[\sin^6(\xi^1)\cos(\xi^1)^{2/3}+D^2]}\;.\ee Then the force in $r$
direction at the highest point is given by \bea\begin{split} \mathscr{L}-\xi'^1{\p\mathscr{L}\ov \p\xi'^1}
=\sqrt{\cos(\xi^1)^{-2/3}[\sin^6(\xi^1)\cos(\xi^1)^{2/3}+D^2]}\;.\end{split}\eea

\begin{figure}[t] \centering
  \includegraphics*[width=0.6\columnwidth]{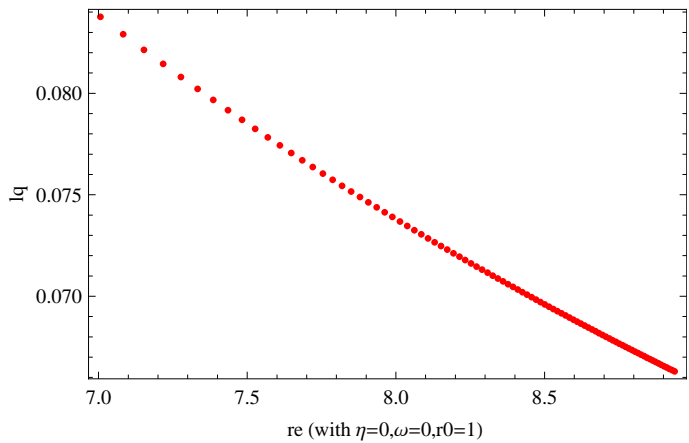}
  \caption{\small $r_e$ dependence of $l_q$ in AdS$_6\times$S$^4$ background.}\label{ads_lq_re}
\end{figure}

The FBC supplies the constraint between $r_e$ and $\rho'(r_e)$. For a given $r(\xi^1=0)$, we can obtain a maximum value
of $r$ in a vertex brane solution, then we can get the string solutions. By analyzing these string solutions in this
AdS$_6\times$S$^4$ background, we plot the curve of $r_e$ dependence of $l_q$ by numerical calculation in Figure
\ref{ads_lq_re}.
We note that there is no maximal value of boundary quark separation. There are two reasons for this phenomenon: one is
that hanging strings in this AdS$_6\times$S$^4$ background do not have similar behaviors as in AdS$_5\times$S$^5$;  the
other is that spike of vertex brane eliminates the critical behavior of hanging strings. We note that the spike
solution of the vertex D4 brane tries to replace the hanging strings. This is a signal to show that, in this
configuration, the vertex brane dominates the properties of baryons.

In this case, we should choose another way to define a ``critical length''(which can be used to judge whether the
baryon is physical, but very different from the usual screening length). We note in the QCD$_4$ deconfined phase,
baryon vertex solutions always touch the boundary brane directly and can not form closed ones, thus we argue that this
touching process corresponds to baryon dissociation in deconfined medium.\footnote{Usually, we know that there is no
baryon states in deconfined phase, since quarks and gluons are deconfined. From the holographic solutions, we see that
solutions in deconfined phase are always unclosed. Thus we think that even if quarks can form a state with the unclosed
solution, it is not a baryon.} From this point of view, we can consider the distance between the top of the vertex
brane and the bottom of the boundary brane as another parameter to judge whether baryons dissociate.

In the AdS$_6\times$S$^4$ background, through analyzing the solutions of D4 vertex brane, we indeed find a maximal
value of $r_{top}$ ( $r_{top}=r_e$ is the radial top position of each solution $r(\theta)$ ) when we change initial
conditions $r(\xi^1=0)$.\footnote{We ignore solutions with $r_{top}>\max\{r_{top}\}$, because there is no spike in
these solutions. This seems just the result from the numerical calculation, we have not found some physical reasons.}
Thus, we can define the maximal top position $l_{m}=\max\{r_{top}\}$ as a critical position. And the critical distance
between $l_m$ and boundary is $l_c=r_\Lambda-l_m$. We can use this critical distance to judge whether a baryon state
can exist due to the following reason. Among these vertex solutions with a spike in AdS$_6\times$S$^4$ background, we
can see from Figure \ref{ads_lq_re} that sizes of baryons are always very small compared with Figure \ref{lq_re}.
Baryons in this background almost stand in a spatial point because the spikes of vertex brane replace hanging strings
to some extent. In this case, the vertex brane dominates main properties of baryons. We can use baryon energy to judge
whether it is physical or not physical. From numerical results, we find that larger $r_{top}$ corresponds to larger
baryon energy, and a solution with $l_m$ has the largest energy. Beyond this $l_m$, there is no solution with a spike,
so we discard those solutions here. Thus $l_m$ gives a upper limit of baryon energy, which is like the usual screening
analysis of holographic baryons. We can understand this point from the rough relation $E_m\sim {1\ov r_\Lambda-l_m}$,
where $E_m$ is the maximal energy of baryons. To avoid possible confusion, we wish to emphasize that $l_m$ or $l_c$ are
not lengths in the gauge theory, but radial parameters in the gravity dual.

Our investigation shows that vertex branes dominate properties of baryons in these configurations. So the velocity
dependence of baryon screening should be obtained by studying on the vertex brane, and high spin baryon should be
described by the vertex brane with inner $J$ charge.
\section{Conclusion and discussion}
How to understand the confinement and calculate hadron spectrum are considered as two biggest problems in QCD(or
non-perturbative QCD exactly). So far we know little about the non-perturbative world and have almost no general
powerful tool to study the strongly coupling problem. AdS/CFT correspondence, which is usually called gauge/gravity
duality in general, is believed as a useful framework to study these problems. In the experiment side, many people
believe there exists a QGP(quark gluon plasma) state in RHIC, which is a strongly coupled quark and gluon thermal state
like a fluid, investigated in many
works~\cite{Dusling:2008tg}~\cite{Erdmenger:2008rm}~\cite{Cai:2008in}~\cite{Cai:2008ph}. How to describe this QGP and
understand the strongly coupled behavior is still a problem, though it is very useful for solving confinement and
hadron spectrum problem. It is believed that heavy quark bound state can be alive in QGP, including J/$\psi$ meson and
some multi-quark bound states. We call these multi-quark bound states baryons, though they may be different from
baryons we see when they survive within QGP. Using meson or baryon as a probe is the simplest method to study the
properties of the strongly coupled quark gluon state.

In the gauge/gravity duality framework, we calculate properties of the probe in the classic gravity background. From
the strong/weak duality, we know these results are always suitable for the probe in the strongly coupled background in
the field side. A lot of works have been done on the meson spectrum and meson melting process in different
gauge/gravity systems. When we consider baryon in the present work, we find following interesting results:

1)In D4-D8 system, a holographic probe baryon can be described as
N$_c$ fundamental strings connect through a vertex D4 brane wrapped
on S$^4$.

 2) In QCD$_4$ background, a closed vertex can exist in confined phase and can not exist in
deconfined phase. In the low temperature region, screening effect
still exist in confined phase like meson and the vertex D4 brane
dominates the baryon mass. The lower energy state corresponds to
vertex brane closer to cut off position ($r=r_c$) and the higher
energy state corresponds to vertex brane closer to boundary (or
flavor brane exactly). We think it is reasonable, because the high
energy limit of this configuration is just like the unclosed vertex
brane configuration in a higher temperature deconfined phase.

 3) In warped
AdS$_6\times$S$^4$ background, a closed vertex can exist in deconfined phase and the vertex contains a spike, and
fundamental strings are relatively short. Screening length should be defined through the distance between top position
of the vertex spike and the boundary.

Related extended works can be done in the future. One is finding more evidence from the experiment data to support live
multi quark bound state in quark gluon plasma. Another is calculating some special parameters of QGP through baryon
probe and comparing them with experiment data. A clear picture of baryon melting is needed and energy loss of baryon
probe is also a very interesting unsolved
problem~\cite{Peigne:2008wu,Marquet:2008kr,Arnold:2008vd,Chesler:2008uy,Peigne:2008nd}.

\section*{Acknowledgements} The author acknowledges helpful discussions with Miao Li, Chaojun Feng, Yi
Wang, Chunshan Lin and thanks Xian Gao for kind help. He also would
like to thank Shi Pu for discussions about the numerical
calculation. In particular, he is very grateful to H.Lu for
suggesting the AdS$_6\times$S$^4$ background and very useful
discussions.

\section*{Appendix}
We denote coordinates in rest frame of quarks as ($t',x_3'$), then we have \bea
\begin{split}
dt&=dt'\cosh\eta-dx_3'\sinh\eta\;,\\
dx_3&=-dt'\sinh\eta+dx_3'\cosh\eta\;. \end{split}\eea The boosted metric is given by \be
ds^2=(\cos\theta)^{-1/3}[Adt^2+2Bdtdx_3+Cdx_3^2+{dr^2 \ov (g^2r^2-{\mu \ov r^3})}+r^2(d\rho^2+\rho^2d\theta^2)] \ee
where \bea
\begin{split}
A&=-(g^2r^2-r^2-{\mu\ov r^3})\cosh^2\eta-r^2\;;\\
B&=(g^2r^2-r^2-{\mu\ov r^3})\;;\\
C&=-(g^2r^2-r^2-{\mu\ov r^3})\sinh^2\eta+r^2\;;
\end{split}
\eea In the warped AdS$_6\times$S$^4$ background, $h_{ab}$ can be read from the induced metric \be
ds_{\tau,\sigma}=(\cos\theta)^{-1/3}[ (A+\sigma^2\rho^2\omega^2)d\tau^2+2Bx_3'd\tau d\sigma+(Cx_3'^2+{1\ov
g^3\sigma^2-{\mu\ov \sigma^3}}+\sigma^2\rho'^2)d\sigma^2]\;, \ee
The string Lagrangian density\be \mathcal
{L}_{string}=(\cos\theta)^{-1/3}\sqrt{(A+\sigma^2\rho^2\omega^2)(Cx_3'^2+{1\ov g^2\sigma^2-{\mu\ov
\sigma^3}}+\sigma^2\rho'^2)-B^2x_3'^2}\;. \ee When the wind velocity $v=0$($\eta=0$), then $x_3'(\sigma)=0$. The
Lagrangian becomes \be \mathcal{L}_{\eta=0}=(\cos\theta)^{-1/3}\sqrt{(-g^2\sigma^2+{\mu\ov
\sigma^3}+\sigma^2\rho^2\omega^2)({1\ov g^2\sigma^2-{\mu\ov \sigma^3}}+\sigma^2\rho'^2)}\;. \ee When angular velocity
$\omega=0$, the Lagrangian becomes \be \mathcal {L}_{\omega=0}=(\cos\theta)^{-1/3}\sqrt{({1\ov g^2\sigma^2-{\mu\ov
\sigma^3}}+\sigma^2\rho'^2)+(AC-B^2)x_3'^2}\;. \ee The EOM of x$_3$: \be (\cos\theta)^{1/3}{\p\mathcal{L}_{string}\ov
\p x'_3}={[(A+\sigma^2\rho^2\omega^2)C-B^2]x'_3\ov \sqrt{(A+\sigma^2\rho^2\omega^2)(Cx_3'^2+{1\ov g^2\sigma^2-{\mu\ov
\sigma^3}}+\sigma^2\rho'^2)-B^2x_3'^2}} =F\;,\ee where $F$ is a constant. The new Lagrangian containing no
$x_3'(\sigma)$ is given by \be
 L=\sqrt{{F^2({1\ov
g^2\sigma^2-{\mu\ov \sigma^3}}+\sigma^2\rho'^2)\ov
(A+\sigma^2\rho^2\omega^2)C-B^2-F^2}+(A+\sigma^2\rho^2\omega^2)({1\ov g^2\sigma^2-{\mu\ov \sigma^3}}+\sigma^2\rho'^2)}
\;.\ee
 All hanging strings attach to the highest point , then the FBC turns to be \be
 L -  \rho' {\p L \ov \p \rho'}\biggr|_{r_e}={2\pi\alpha'(\cos\theta)^{1/3}\ov N_c}
 {\p \mathcal {H} \ov \p r_e} \;,
\ee where $r_e=r_{max}$(with $r'(\xi^1)=0$) in the brane solution. To calculate the ${\p \mathcal {H} \ov \p r_e}$, we
rewrite the new Lagrangian of D4 brane as \be \begin{split}\mathcal {H} &= -T_4\Omega_3g_s({\sqrt{2}\ov g})^3\int
d\xi^1\sqrt{[r'(\xi^1)+2g^{-2}f(r)]\cos(\xi^1)^{-2/3}[\sin^6(\xi^1)\cos(\xi^1)^{2/3}+D^2]}\\
&=-T_4\Omega_3g_s({\sqrt{2}\ov g})^3\int
dr\sqrt{[1+2g^{-2}f(r)\xi'^1(r)]\cos(\xi^1)^{-2/3}[\sin^6(\xi^1)\cos(\xi^1)^{2/3}+D^2]}\;.
\end{split}
\ee We denote \be \mathscr{L}=
\sqrt{[1+2g^{-2}f(r)\xi'^1(r)]\cos(\xi^1)^{-2/3}[\sin^6(\xi^1)\cos(\xi^1)^{2/3}+D^2]}\;.\footnote{$\xi'^1(r)={\p\xi^1\ov
\p r.}$}\ee Then the force in $r$ direction at the highest point is given by \bea\begin{split}
\mathscr{L}-\xi'^1{\p\mathscr{L}\ov \p\xi'^1} &={\sqrt{\cos(\xi^1)^{-2/3}[\sin^6(\xi^1)\cos(\xi^1)^{2/3}+D^2]}\ov
\sqrt{1+2g^{-2}f\xi'^1}}\mid_{\xi'^1=0}\\
&=\sqrt{\cos(\xi^1)^{-2/3}[\sin^6(\xi^1)\cos(\xi^1)^{2/3}+D^2]}\;.\end{split}\eea

\end{document}